\documentclass[usenatbib,usegraphicx,times]{mn2e}

\newcommand{\eb}{\begin{equation}}
\newcommand{\ee}{\end{equation}}
\title[Forced librations]{Forced libration of tidally synchronized planets and moons}
\author[Valeri V. Makarov et al.]{Valeri V. Makarov$^{1}$\thanks{E-mail: vvm@usno.navy.mil}, Julien Frouard$^{1,2}$ 
\& Bryan Dorland$^{1}$
\\
$^{1}$US Naval Observatory, 3450 Massachusetts Ave NW, Washington DC 20392-5420, USA\\
$^2$ Federated IT}
\begin{document}

\date{Accepted . Received ; in original form }

\pagerange{\pageref{firstpage}--\pageref{lastpage}} \pubyear{2015}

\maketitle

\label{firstpage}

\begin{abstract}
Tidal dissipation of kinetic energy, when it is strong enough, tends to synchronize the rotation of planets and moons with
the mean orbital motion, or drive it into long-term stable spin-orbit resonances. As the orbital motion undergoes
periodic acceleration due to a finite orbital eccentricity, the spin rate oscillates around the equilibrium mean value
too, giving rise to the forced, or eccentricity-driven, librations. Both the shape and amplitude of
forced librations of synchronous viscoelastic planets and moons are defined by a combination of two different types
of perturbative torque, the tidal torque and the triaxial torque. Consequently, forced librations can be tidally dominated
(e.g., Io and possibly Titan) or deformation-dominated (e.g., the Moon) depending on a set of orbital, rheological,
and other physical parameters. With small eccentricities, for the former kind, the largest term in the libration 
angle can be minus cosine of the mean
anomaly, whereas for the latter kind, it is minus sine of the mean anomaly. The shape and the amplitude of tidal forced
librations determine the rate of orbital evolution of synchronous planets and moons, i.e., the rate
of dissipative damping of semimajor axis and eccentricity. The known super-Earth exoplanets
can exhibit both kinds of libration, or a mixture thereof, depending on, for example, the effective Maxwell time of their rigid mantles. Our approach
can be extended to estimate the amplitudes of other libration harmonics, as well as the forced libration in
non-synchronous spin-orbit resonances. \end{abstract}


\section{Introduction}
Most of the moons and satellites in the Solar system have a small orbital eccentricity and are locked in
the 1:1 spin-orbit resonance, when the period of sidereal rotation is exactly equal to the orbital period.
The commonly accepted explanation for both phenomena is sufficiently strong tidal interactions with their host
planets, which whittle down the angular momentum and kinetic energy over the course of billions of years
and drive the system into the global potential well. Mercury represents a notable exception with its relatively
high eccentricity and a stable 3:2 spin-orbit resonance. Tidal dissipation of orbital energy continues in
resonantly rotating bodies as long as the eccentricity is nonzero. An analytical solution for energy dissipation
in a closed form was obtained for the simplest model case of a uniform (unstructured) spherical body
\citep[][and references therein]{peca,heat1}.  This analysis implicitly assumes that the spin rate of the body
is nearly constant over one orbit, leading to a zero contribution from the leading (resonant) tidal mode.
In reality though, the spin rate of a resonant body is subject to periodic oscillations, called librations in
longitude. In most cases, librations are caused by the strictly periodic torque acting on the perturbed body
(i.e., the body with tides) resulting from the misalignment of the ``long" axis of the smallest moment of inertia
and the instantaneous direction to the center of the perturbing body. The misalignment comes from the periodic
acceleration of the perturber on the eccentric orbit. Bodies with complete rotational symmetry about the principal
axis of inertia, such as completely liquid planets and stars, do not have a permanent figure and therefore,
are not subject to this type of libration. Since the main type of libration is related to a certain force (in this
case, the force of gravitation from the perturber acting on the triaxial figure), it is called {\it forced}
libration.

\citet{wis04} suggested that forced librations of resonantly rotating bodies boost the dissipation of orbital
energy and heating of the perturbed body. This possibility was also discussed by \citet{me14} in the
framework of a general viscoelastic tidal model. The boost may happen because the secular component of the tidal torque
depends on the instantaneous spin rate, rather than on the instantaneous libration angle, and may result in
a negative work, being in opposite-phase with the velocity. Physically, the tidal bulge is in a small-angle
periodical motion across the surface of the body with a certain phase lag, giving rise to additional internal
friction. This surmise lacks a detailed, quantitative analysis. Qualitatively, this extra dissipation is expected to
be small for nearly circular orbits and small libration amplitudes (such as the lunar forced libration), and
the traditional approach to tidal heating estimation \citep{peca} is perhaps justified.

Numerous exoplanets have been found close enough to their host stars for the tidal forces to be important in
the dynamical heating and orbital evolution. Close-in super-Earths, in particular, often seem to have unrelaxed eccentricities. They should be locked in the 1:1 or higher order spin-orbit resonances. The vividly discussed
issue of possible habitability of such planets can not be decoupled from spin-orbit dynamics in general and libration
in particular. \citet{dob} considered the important role of {\it optical} librations for climate and habitability of
exoplanets, but ignored the physical forced librations. Optical libration is a merely geometrical effect
related to the changing aspect angle of the satellite surface relative to the line through the centers even
in the absence of spin rate oscillations. Even on synchronized planets, a small amount of physical rocking
motion may generate rather extensive zones of temperate irradiation suitable for biological life. For now, the
parameters of exoplanet libration remain a matter of theoretical speculation, as there are no observational means to
detect or measure it.

Small satellites of strongly asymmetric or deformed shapes represent another interesting area for libration theory applications, this time with a distinct possibility of verification by observational evidence. \citet{tho}
measured a large libration longitude angle of $0.12$ degrees for Enceladus, which they explained by the presence
of a massive subsurface ocean. A large, nearly decoupled molten core in Mercury is betrayed by the observed amplitude of the
forced libration \citep{tay}, which is roughly twice as large as what is expected from a uniformly rigid planet.
A non-vanishing friction between the solid mantle and the liquid core, or between the crust and a subsurface ocean in
case of smaller satellites, can change not only the magnitude but also the spectrum of observed short-period librations.
Additional complications arise for small satellites with a large degree of elongation when the natural oscillation
frequency becomes commensurate with the frequency of forced excitation. Both the capture of such bodies into the spin-orbit
resonance \citep{ale} and the post-capture rotation \citep{cor} may be chaotic. 

\citet{gol} investigated a more complex case of attenuated forced librations in a liquid body covered by a crust.
The dynamic elongation of the liquid component (e.g., a subsurface ocean) interacts with the elastically deformable
crust, the latter resisting the motion of the tidal bulge across the surface, resulting in additional dissipation
of kinetic energy and reduction in the amplitude of libration. This model was further worked out and applied to
realistic icy satellite scenarios by \citet{van}. Another complex model of librations in the three-body Sun-Earth-Moon
system was developed by \citet{wil}. There the authors pointed out that tidal interactions and possible resonances
between them can change the spectrum of observed librations. This elaborate model has been compared with lunar ranging
observations.

In this paper, we consider the simplest case of a settled binary system with tides in the 1:1 (synchronous) resonance
with a small eccentricity. We consider a uniform body neglecting subtle effects that may be caused by a
multi-layered structure. The novelty of our approach is in an accurate analysis of the interplay between the gravitational
and tidal perturbative torques, the latter having been often ignored in the literature on libration. 
We will show that even
within this simple set-up, a new type of forced libration emerges, drastically different from the relatively
well studied model of Lunar and Mercury's libration. 

 \begin{table*}
 \centering
 \caption{Assumed parameters of the Moon, Io, and Kepler-10b.}
 \label{moon.tab}
 \begin{tabular}{@{}lrrrrr@{}}
 \hline
            &                 &       &\\
   Name     &  Description    & Units & Moon & Io & Kepler-10b\\
 \hline
 $R$   & \dotfill radius of perturbed body             & m & $1.373\cdot 10^6$& $1.821\cdot 10^6$& $9.01\cdot 10^6$\\
 $M_2$ & \dotfill mass of the perturbed body & kg & $7.3477\cdot 10^{22}$& $8.932\cdot 10^{22}$& $2.65\cdot 10^{25}$\\
 $M_1$ & \dotfill mass of the perturbing body& kg & $5.97\cdot 10^{24}$& $1.899\cdot 10^{27}$& $1.78\cdot 10^{30}$ \\
 $a$ & \dotfill semimajor axis & m & $3.844\cdot 10^{8}$& $4.217\cdot 10^{8}$& $2.5\cdot 10^{9}$\\
 $n$ & \dotfill mean motion, i.e. $2\pi/P_{\rm orb}$ & yr$^{-1}$ & 84 & 1297& 2742\\
 $e$ & \dotfill orbital eccentricity & & 0.0549& 0.0041& 0.01\\
 $(B-A)/C$ & \dotfill triaxiality &  & $2.278\cdot 10^{-4}$ & $6.4\cdot 10^{-3}$& $2.03\cdot 10^{-4}$ \\
 $\tau_M$ & \dotfill Maxwell time  & var.  &  11 yr& 10 days& 1 days\\
 $\mu$ & \dotfill unrelaxed rigidity modulus & Pa
   & $0.65\cdot10^{11}$& $0.65\cdot10^{11}$& $0.8\cdot10^{11}$\\
 \hline
 \end{tabular}
 \end{table*}

\section{Equations of periodic libration}
Ultimately, the perturbing torque generating forced librations is related to the orbital acceleration of the
perturber, which is strictly periodic with the orbital period as long as long-term orbital evolution or precession of the equator is
ignored. The terms of forced libration in this simplified setup are harmonics of the main orbital frequency.
We can therefore represent the libration angle in the general form as a Fourier series of the mean anomaly ${\cal M}$:
\eb
\theta-{\cal M}=\sum_{s=1}^\infty \alpha_s\sin\,s{\cal M}+\sum_{p=1}^\infty\beta_p\cos\,p{\cal M}
\label{lib.eq}
\ee
with the coefficients $\alpha_s$ and $\beta_p$ to be determined from the equation of forces. The constant term
$p=0$ is omitted in this expansion because its value is dependent on the choice of the origin for the
angular coordinate $\theta$. It is practical to reckon $\theta$ from the direction to the pericenter of the perturber
as seen from the perturbed body to the instantaneous direction of the longer principal axis associated with the
smallest moment of inertia $A$ (let us call it ``$A$-axis"). Still, 
the action of tidal torque on a triaxial body tends to accelerate the
prograde rotation of a synchronized body (in other words, the time-average tidal torque is positive). The
equilibrium is achieved through a negative time-average triaxial torque in the counter direction, which requires 
the $A$-axis to be leading the mean direction of the perturber \citep{mamoon}. For the Moon, this constant tilt amounts to
67.753 arcsec \citep{ram}. Being in itself an important observable parameter, this tilt does not contribute
to the energy dissipation or orbital evolution.

Differentiating Eq. \ref{lib.eq} in time and using $\dot{\cal M}=n$ obtains
\begin{eqnarray}
\dot\theta-n&=& n\left[ \sum_{s=1}^\infty \alpha_s\,s\cos\,s{\cal M}-\sum_{p=1}^\infty\beta_p\,p\sin\,p{\cal M}\right]
 \nonumber \\
\ddot\theta&=& -n^2\left[ \sum_{s=1}^\infty \alpha_s\,s^2\sin\,s{\cal M}+\sum_{p=1}^\infty\beta_p\,p^2\cos\,p{\cal M}\right].
\label{lib2.eq}
\end{eqnarray}

The equation of forces is
\eb
\xi M_2 R^2\ddot\theta=T_{\rm TRI}+T_{\rm TIDE}
\ee
with $T_{\rm TRI}$ and $T_{\rm TIDE}$ being the triaxial and tidal torques, respectively, $M_2$ the mass of the
perturbed body, $R$ its radius, and $\xi$ the moment of inertia coefficient usually ranging between 0.3 and 0.4.

\section{Components of tidal torque}
We are using equations for the polar tidal torque (i.e., the component directed along the axis of rotation) from \citep{ma12}. Those expressions are based on the theory of tidal torque developed
by \citet{efr1,efr2} for a homogeneous near-spherical body of an essentially arbitrary rheology. As in {\it{ibid.}}, 
we chose a realistic rheological law combining viscoelastic (Maxwell) response
and inelastic (Andrade) creep caused by defect unpinning. Both components of the model
are strongly frequency-dependent in the vicinity of spin-orbit resonances, and they can not be decoupled in
a linear approximation. The inherent nonlinearity of the functional dependence on frequency in this model leaves 
little room for analytical applications, compared with
the commonly used Constant Time Lag (CTL) model. However, we have to exploit more complex and accurate models
in specific applications rather than the CTL model, which can not correctly explain the spectrum of lunar
libration, for example, or the rate of eccentricity change in the Earth-Moon system. Among more advanced tidal
models, \citet{fm} proposed a theoretical framework emphasizing the role of the Andrade dislocation and creep.
Here, on the contrary, we will exploit a purely Maxwell approximation, i.e., a uniform viscoelastic body with
rigidity and self-gravitation. Such approximations may shed light on the tidal behavior of planets and satellites
of relatively low effective viscosity, for example, with post-solidus mantles including partial melt.
 
As was first pointed out by \citet{efr1}, the tidal torque includes both secular and oscillating components of
time (or mean anomaly ${\cal M}$):
\begin{eqnarray}
\ddot\theta_{\rm TIDE}&=&\frac{3}{2}\frac{M_1}{\xi M_2}n^2\left(\frac{R}{a}\right)^3\sum_{q} G_{20q}(e)
\sum_{j} G_{20j}(e) \nonumber\\  & &
[K_c(2,\nu_{220q})\,{\rm Sign}(\omega_{220q})\cos\left((q-j){\cal M}\right) -\nonumber \\
&& K_s(2,\nu_{220q})\sin\left((q-j){\cal M}\right)],
\label{tide.eq}
\end{eqnarray}
where $\xi$  is the coefficient of inertia, $C/(M_2R^2)$, $G_{lpq}$ are Kaula's functions of eccentricity related 
to Hansen's coefficients through $G_{20(j-2)}=X^{-3,\,2}_j$, the tidal mode $\omega_{220q}=(2+q)n-2\dot\theta$,
and the quality functions $K_c$ and $K_s$ are
described in \citep{mak15}. The secular terms can be seen emerging from the overlap
of tidal modes, i.e., $q=j$. In Section \ref{tide.sec}, we shall explain why the oscillating terms $q\neq j$ can not be always ignored in the consideration of
forced libration.

\section{Triaxiality-driven libration}
\label{3.sec}
When the mass of the perturbed body is much smaller than the mass of the (primary) perturber, $M_2 \ll M_1$, the
instantaneous acceleration of the deformed, triaxial body can be written as \citep[e.g.,][]{noy}
\eb
\ddot\theta_{\rm TRI}=-\frac{3}{2}\frac{B-A}{C}n^2\sum_j G_{20(j-2)}\sin(2\theta-j {\cal M}),
\label{tri.eq}
\ee
Note that the summation in Eq. \ref{tri.eq} is over $j=-\infty,\ldots,+\infty$. The three principal moments of inertia $A$, $B$, and $C$ are unequal, with $C>B>A$ by definition. The angle of rotation $\theta$ in this equation is not a linear
function of time (or $\cal{M}$), but varies periodically for a synchronous body in accordance with Eq. \ref{lib.eq}.
Rigorous substitution of the Fourier series into Eq. \ref{tri.eq} results in intractable infinite trigonometric series 
with the Bessel function coefficients. Fortunately, for the most important case of small librations, when
$\theta-{\cal M}$ is small, we can write, approximately,
\eb
\ddot\theta_{\rm TRI}\approx Zn^2\sum_{p=1}^{+\infty}(G_{20p}-G_{20(-p)})\sin{p\cal M},
\label{3.eq}
\ee
where 
\eb
Z=\frac{3}{2}\;(B-A)/C.
\ee
 Here we also limit our analysis to the case of slightly
deformed bodies with $(B-A)/C\ll 1$. Also in the following, we will only consider
the main term proportional to $\sin{\cal M}$. This is justified as long as the orbital eccentricity is small,
because the amplitude of $j\neq 2$ terms is approximately proportional to $G_{20(j-2)}$, the most significant
Kaula's functions being $G_{20(-1)}=-1/2\;e+O(e^3)$, $G_{200}=1-5\;e^2+O(e^4)$, $G_{201}=7/2\;e+O(e^3)$, $G_{202}=17/2\;e^2+O(e^4)$. Keeping only the series terms up to $O(e)$, one immediately obtains
\eb
\ddot\theta_{\rm TRI}\approx 4Zn^2e\sin{\cal M}.
\label{ec.eq}
\ee
For a derivation of the complete Fourier series of triaxiality-driven libration see \citep{com}.

In the absence of other torques, an estimate for the main component of forced libration is readily obtained from
Eq. \ref{ec.eq}, 
\eb
\alpha_1\approx -4Ze.
\ee
This formula was derived by \citet{eck}. 

This simple calculation provides a coefficient of $-15.5$ arcsec with our best knowledge physical parameters for the
Moon listed in Table \ref{moon.tab}. We find close agreement with the observed values from \citep{ram}, as well as
with our own more accurate computation including the tidal torque, see Table \ref{harm.tab}. The tides on the Moon
raised by the Earth result in the emergence of cosine harmonics of libration, most notably, of a positive (leading)
tilt constant in time, but they little change the amplitudes of the sine harmonics. In this respect, the influence
of the tidal torque on the Moon is small. Why?

 \begin{table*}
 \centering
 \caption{Coefficients of forced libration harmonics of the Moon, in arcseconds.}
 \label{harm.tab}
 \begin{tabular}{@{}lcr@{}}
 \hline
            &                 &       \\
   Term     &  \citet{ram}    & Our calculation\\
 \hline
 $\cos0{\cal M}$   & $+67.753$  & $+66.9$ \\
 $\cos1{\cal M}$ & $-0.004$ & $-1.617$\\
 $\sin1{\cal M}$ & $-16.799$ & $-15.28$\\
 $\sin2{\cal M}$ & $-0.445$ & $-0.442$ \\
 $\sin3{\cal M}$ & $-0.022$ & $-0.064$\\
 \hline
 \end{tabular}
 \end{table*}

\begin{figure*}
\begin{minipage}{82mm}
\includegraphics[width=82mm]{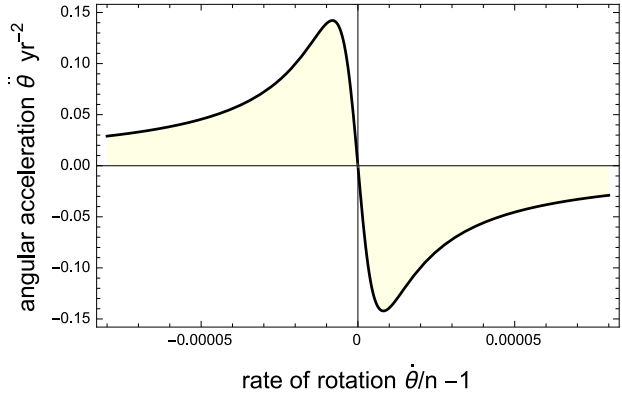}
\end{minipage}\hspace{2pc}
\caption{Angular acceleration of the Moon caused by the secular component of the tidal torque as a function of normalized
relative spin rate, computed for a Maxwell time equal to 11 years. The span of this graph in the horizontal axis ($\dot\theta/n-1$)
is approximately equal to the full amplitude of the Moon's forced libration in angular velocity. Observe that the peaks of
tidal dissipation are much closer to the resonance than the extent of libration, effectively diminishing the impact of the dissipative term.}
\label{kvalitet.fig}
\end{figure*}

The secular tidal torque in the viscoelastic Maxwell model with self-gravitation considered in this paper is a strongly nonlinear function of
tidal mode in the vicinity of the 1:1 spin-orbit resonance (Fig. \ref{kvalitet.fig}). 
As explained in detail in \citep{mabe},
this function can be broken into two components, an antisymmetric kink shape and a nearly constant bias term,
defined by the contribution of non-resonant tidal modes. The bias, or the offset, is barely noticeable in
Fig. \ref{kvalitet.fig}, but it does shift the function up by $2.2\cdot 10^{-5}$ yr$^{-2}$ at zero frequency.
The secular prograde acceleration coming from the tides has to be counterbalanced by a secular triaxial torque
acting in the opposite direction. The balance is automatically achieved and maintained, as the accelerating
rotation increases the lead of the long axis relative to the average direction to the perturber until the perfect equilibrium of torques is reached. The bias thus generates the average tilt, i.e., the $\cos\,0{\cal M}$ harmonic
of libration. The amplitude of this tilt is especially sensitive to the Maxwell time parameter of the model, which helps
to constrain it to $\approx 11$ yr for the Moon. The kink is characterized by a couple of opposed peaks and a
roughly linear segment between them. Within this segment, and only there, this model becomes very similar
to the much-explored Constant Time Lag model of tides.

The peaks are separated from the point of resonance ($\omega_{2200}=0$, where $\omega_{220q}=(2+q)n-2\dot\theta$) by approximately \citep{mak15,efr15}
\eb
(\omega)_{\rm peak}=\frac{1}{\tau_M(1+\Lambda_2)}.
\label{peak.eq}
\ee
Note that the main tidal mode and frequency are {\it semidiurnal} for the synchronous rotation, i.e.,
$|\omega_{2200}|=\nu_{2200}=|2(n-\dot\theta)|$.
The dimensionless parameter $\Lambda_2$, often called effective rigidity and denominated $A_2$ in the literature,
emerges from the definition of the quadrupole static Love number \citep{mac}. The peak tidal quality at this wave number is
\eb
K_c^{\rm peak}=\frac{3}{4}\frac{\Lambda_2}{1+\Lambda_2}.
\label{kpeak.eq}
\ee
The loci of the peak torque depend on both the Maxwell time and effective rigidity $\Lambda_2$, whereas the magnitude
depends only on the latter. The effective rigidity varies in a relatively narrow range between $\sim1$ for
super-Earth exoplanets, which are possibly the most massive planets of terrestrial composition, and several tens
for the smaller satellites of the Solar system (for the Moon, it is approximately equal to 65). The range of $\tau_M$ is many orders of magnitude, because the
viscosity of silicate minerals is a steep function of temperature \citep[e.g.,][]{beh}. Therefore, the
spread of the tidal peaks near a resonance is quite sensitive to the effective viscosity of the dissipating layers.
Cold, inviscid planets such as Earth and the Moon have Maxwell times of tens to hundreds of years. Semiliquid or
icy satellites such as Io and Titan are likely to have this parameter in the order of days. Consequently, the peaks
can be tightly packed at the point of resonance, as we see for the Moon in Fig. \ref{kvalitet.fig}, or they can spread
out as far as the adjacent resonances, or even farther. 

The forced libration remain triaxiality-driven when the peaks are well within the range of regular velocity oscillation.
Combining Eqs. \ref{ec.eq} and \ref{peak.eq}, the resulting condition for synchronous rotation is
\eb
\tau_M\gg \left(8\;Zen(1+\Lambda_2)\right)^{-1}.
\label{zen.eq}
\ee
When this condition is fulfilled, the $\sin{\cal M}$ variation of velocity takes the tidal torque well over the
peak values to the domain where it becomes rather insignificant. This happens because the perturbed body, rocked by
the periodically accelerating perturber, passes the area of tidal peaks very quickly at ${\cal M} \approx 0$ and $\pi$ (i.e., at the pericenter and apocenter), spending more time in quadratures, where the spin rate deviates the most from
the resonance and the tidal torque is weak. The critical value of $\tau_M$ for the Moon beyond which the forced libration
is triaxiality-driven, is 1.2 yr.

\begin{figure*}
\begin{minipage}{82mm}
\includegraphics[width=82mm]{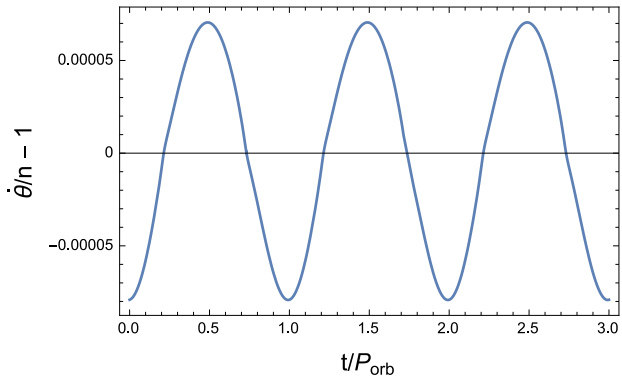}
\end{minipage}\hspace{2pc}
\caption{Numerically simulated libration of the Moon in relative normalized
spin rate, $(\dot\theta-n)/n$, for three complete orbital
periods.}
\label{moon.fig}
\end{figure*}

\begin{figure*}
\begin{minipage}{82mm}
\includegraphics[width=82mm]{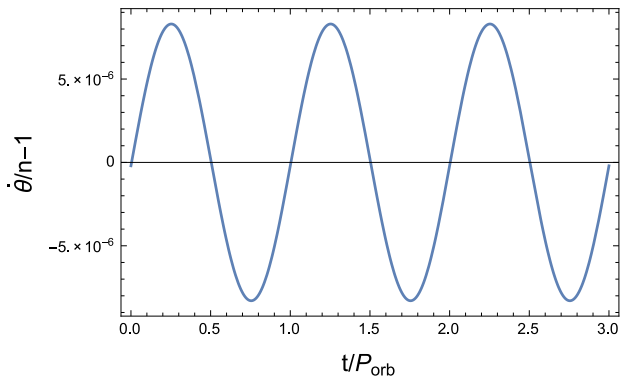}
\end{minipage}\hspace{2pc}
\caption{Numerically simulated libration of Io in spin rate, $\dot\theta-n$, for three complete orbital
periods.}
\label{io.fig}
\end{figure*}

\section{Libration of highly elongated synchronous bodies}

Equation \ref{ec.eq} is accurate enough for rotating nearly symmetric bodies with small eccentricities,
but smaller satellites often have large triaxiality parameters $(B-A)/C$. A more accurate treatment may
be required, taking into account the possible interferences with the natural mode of free libration.
The classical equation of the spin-orbit problem with a trixial torque reads \citep{wis,mur}
\begin{equation}
\ddot{\theta} = \frac{3}{2} \frac{B-A}{C} \frac{G m_p}{a^3} \frac{a^3}{r^3} \sin 2 (f-\theta).
\end{equation}
To compute the evolution of the libration angle in the synchronous resonance, we make the change of variables $\gamma = \theta - pM$, choose $p=1$ to select the synchronous resonance, and get
\begin{equation}
\ddot{\gamma} =  \frac{\omega_0^2}{2} \frac{a^3}{r^3} \sin 2 (f-M-\gamma) 
\label{julieneq1}
\end{equation}
where $\omega_0^2 = 3 \frac{B-A}{C} \frac{G m_p}{a^3}$. If we choose to neglect $\gamma$ on the r.h.s. of the above equation (because of its smallness close to the exact resonance), we can expand $\frac{a^3}{r^3} \sin 2 (f-M)$ in eccentricity and obtain the Fourier decomposition of $\gamma(t)$ in multiples of $nt$ (see the complete Fourier decomposition in \cite{com}). On the other hand, we can obtain a more complete formulation in case $\omega_0$ is not negligible compared to $n$, by expanding Eq.\ref{julieneq1} in Taylor series of $\gamma$ and keeping only the terms of zeroth and first order: 
\begin{equation}
\ddot{\gamma} = \frac{\omega_0^2}{2} \frac{a^3}{r^3} \bigg[ \sin 2 (f-M) - 2 \gamma   \cos 2 (f-M) \bigg].
\end{equation}
Expanding this equation in terms of the eccentricity functions $F_q^+,F_q^-$ of \citep{com}\footnote{Mind the small misprint in Table 2 of Comstock \& Bills (2003) where $F_1^- = G_{2,0,-1} = -\frac{e}{2} + \frac{e^3}{16} - \frac{5 e^5}{384}$}, we have
\begin{equation}
\ddot{\gamma} + \omega_0^2 \gamma = - \frac{\omega_0^2}{2} \sum_{q=1} F_q^+ \sin [(2+q)nt] + F_q^- \sin [(2-q)nt]
\end{equation}
where we kept only the zeroth order term in the expansion of $\frac{a^3}{r^3} \cos 2 (f-M) = 1 + 3 e\cos M + e^2 (- \frac{5}{2} + \frac{17}{2} \cos 2M) + \mathcal{O}(e^3)$. This equation can be solved to give the Fourier decomposition of the libration angle
\begin{eqnarray}
\gamma(t) = & A_0 \cos(\omega_0 t + \phi_0)  \nonumber\\
	    &+ ( \frac{\omega_0}{2} )^2 \sum_{q=1} \frac{C_q^{-3,2} - S_q^{-3,2} }{n^2 (2+q)^2 - \omega_0^2} \sin[(2+q)nt] 
	    \nonumber\\
	    &+ \frac{C_q^{-3,2} + S_q^{-3,2} }{n^2 (2-q)^2 - \omega_0^2} \sin[(2-q)nt]
\end{eqnarray}
where the $C_q^{-3,2},S_q^{-3,2}$ are the Cayley coefficients (see \cite{com}) and $A_0,\phi_0$ depend on initial conditions. We can merge terms with similar frequencies to obtain
\begin{eqnarray}
\gamma(t) = & A_0 \cos(\omega_0 t + \phi_0)  \nonumber\\
	    &  + ( \frac{\omega_0}{2} )^2 \frac{C_1^{-3,2} + S_1^{-3,2}  - C_3^{-3,2} - S_3^{-3,2}}{n^2 - \omega_0^2} \sin(nt) \nonumber\\
	    &  - ( \frac{\omega_0}{2} )^2 \frac{(C_4^{-3,2} + S_4^{-3,2})}{4 n^2 - \omega_0^2} \sin(2nt) \nonumber\\    
	    &+ ( \frac{\omega_0}{2} )^2 \sum_{q=3} \frac{C_{q-2}^{-3,2} - S_{q-2}^{-3,2} - C_{q+2}^{-3,2} - S_{q+2}^{-3,2}}{q^2 n^2  - \omega_0^2} \sin(qnt)
\label{sin.eq} \end{eqnarray}

The first term of this equation represents the regular free libration at the frequency $\omega_0$ with arbitrary
initial parameters. Free libration is rather quickly damped at resonances and is not considered in this paper.
The second term is a refinement of the main mode of forced libration (\ref{ec.eq}), with a correction coefficient
$n^2/(n^2-\omega_0^2)$. This correction implies that the amplitude of forced libration becomes infinitely large when
$n=\omega_0$, or $(B-A)/C\approx 1/3$ for the perturber's mass $m_p$ much greater than the mass of the perturbed body.
This only means that the above equation does not work at the points of singularity $\omega_0=q\;n$ for positive 
integer $q$ \citep{wis}. Some satellites have triaxiality parameters close to $1/3$, for example, Epimetheus, 
with a $(B-A)/C\approx 0.30$
\citep{tis} or 0.28 \citep{cau}. Should we expect a very large $\sin {\cal M}$ forced libration for Epimetheus in
accordance with Eq. \ref{sin.eq}? More detailed dynamical analysis indicates that instead of singularity and a sudden
change of libration phase, another island of equilibrium emerges in the phase space \citep{mel} with the average
rotational velocity still locked in the 1:1 resonance.

\section{Tidally driven libration}
\label{tide.sec}
We have considered the case of relatively cold and high-viscosity bodies with Maxwell times significantly larger
than the critical value defined in Eq. \ref{zen.eq}. This model can be used for the Moon, Earth and Mercury.
At the other end of the gamut, semiliquid, semi-molten planets or icy satellites are characterized by low
viscosity but finite sheer modulus, resulting in a short Maxwell time well below the critical value.
We will see in the following that this circumstance may lead to a very different spectrum of forced libration.

So, if 
\eb
\tau_M\ll \left(8\;Zen(1+\Lambda_2)\right)^{-1},
\label{cri.eq}
\ee
the ``secular" component of tidal acceleration can be well approximated as
\eb
\ddot\theta_s\approx -Yn(\dot\theta -n),
\label{s.eq}
\ee
where
\eb
Y=\frac{9}{2}\frac{M_1}{\xi M_2}n\tau_M \left(\frac{R}{a}\right)^3\Lambda_2G^2_{200}(e),
\ee
see \citep{mak15}. This approximate equation just reflects the fact that the segment of tidal torque between
the peaks resembles a straight line with negative slope (Fig. \ref{kvalitet.fig}). Since the main term of
libration in angle emerging from the triaxial torque is proportional to $-\sin{\cal M}$, the main term in
spin rate is proportional to $-\cos{\cal M}$, and the acceleration $\ddot\theta_s$, by Eq. \ref{s.eq}, is
proportional to $+\cos{\cal M}$. Therefore, this component of tidal torque gives rise to libration angle
terms proportional to $-\cos{\cal M}$. Being in phase with the triaxiality-driven spin rate oscillation,
it defines the rate of energy dissipation and the orbital evolution of synchronized bodies.

The oscillating part of the tidal torque includes both $\sin j{\cal M}$ and $\cos j{\cal M}$ harmonics \citep{efr2,
mak15}. When the eccentricity of a synchronously rotating body is small, $e<0.1$, and the $\tau_M$ is longer than the
orbital period, the former lot can be
neglected, because the kvalitet function $K_c(2,\omega_{20q})$ is usually very small at non-resonant tidal modes ($q\neq 0$).
This can not be said about the sine-harmonics of the oscillating torque. The corresponding acceleration is,
specifically,
\eb
\ddot\theta_o=-\hat X n^2\sum_q\sum_{j\neq q} G_{20q}(e)G_{20j}(e)K_s(2,\omega_{20q})\sin(q-j){\cal M},
\label{o.eq}
\ee
where 
\eb
\hat X=\frac{3}{2}\frac{M_1}{\xi M_2}\left(\frac{R}{a}\right)^3.
\ee
The kvalitet function is, specifically,
\eb
K_s(\omega)=\frac{3}{2}\frac{(\omega\tau_M)^2(1+\Lambda_2)+1}{(\omega\tau_M)^2(1+\Lambda_2)^2+1},
\ee
where we dropped the indices $20q$ for all frequencies. At zero (resonant) tidal frequency, $K_s(0)=3/2$, which
is at least twice as large than the peak quality for the secular part and sine-harmonics (Eq. \ref{kpeak.eq}).
It follows that the sine terms can be quite significant.

Combining the last of Eqs. \ref{lib2.eq} with \ref{3.eq}, \ref{s.eq}, and \ref{o.eq}, one obtains an infinite
system of equations:
\begin{eqnarray}
\alpha_s s^2 &=& -Z(G_{20s}-G_{20(-s)})-Y\beta_s s+\hat X\sum_q G_{20q}(e) \nonumber\\
& & (G_{20(q-s)}(e)-G_{20(q+s)}(e))K_s(2,\omega_{20q})\\
\beta_p p &=& Y\alpha_p.
\label{sys.eq}
\end{eqnarray}
Let us further restrict our analysis to the case of small eccentricity ($e<0.1$), keeping only terms to $O(1)$
and $O(e)$ in the expansions of the Kaula's function products. This effectively eliminates all $\alpha_s$
unknowns except $\alpha_1$, for which we get

\eb
\alpha_1\approx -4Ze-Y^2\alpha_1+4\hat X e(K_s(2,n)-K_s(2,0)).
\ee
Introduce
\eb
\Delta K=K_s(2,0)-K_s(2,n)=\frac{3}{2}\frac{\lambda^2\Lambda_2(1+\Lambda_2)}{\lambda^2(1+\Lambda_2)^2+1},
\ee
where the tidal wave number $\lambda=\tau_M n$,
leading to an approximate solution for the $\sin{\cal M}$ and $\cos{\cal M}$ terms
\begin{eqnarray}
\alpha_1  &\approx& -4e\frac{Z+\hat X\Delta K}{1+Y^2}\\
\beta_1  &\approx& Y\alpha_1.
\label{ans.eq}
\end{eqnarray}

Two important conclusions for tidally synchronized bodies emerge right away: 1) the $\cos{\cal M}$ and $\sin{\cal M}$
terms in libration angle are both negative, and their ratio is defined by the ``tidal feedback" coefficient
$Y$; 2) The classic, triaxiality driven libration term $-4Ze\sin{\cal M}$ is severely suppressed if $Y\gg 1$,
where the $\cos{\cal M}$ becomes dominant.

\section{Summary and discussion}
\label{discu.sec}
The three dimensionless parameters, $X=\hat X \Delta K$, $Y$, and $Z$, as well as the orbital eccentricity $e$, define
the spectrum of forced libration of tidally synchronized bodies. Even though our knowledge of the physical parameters
listed in Table \ref{moon.tab} is imperfect, especially for the exoplanet Kepler 10b, estimating the $X,Y,Z$
parameters for the three objects allows us to outline the range of possible libration spectra. Table \ref{xyz.tab} gives
our best-guess estimates for the Moon, Io, and Kepler 10b.

 \begin{table*}
 \centering
 \caption{Coefficients $X,Y,Z$ defining the spectrum of forced libration, estimated the Moon, Io, and Kepler-10b.}
 \label{xyz.tab}
 \begin{tabular}{@{}lrrr@{}}
 \hline
    & Moon & Io & Kepler-10b\\
 \hline
 $X$    & $0.00004$& $0.0095$& $0.0054$\\
 $Y$ &  $5.02$& $36.6$& $0.12$\\
 $Z$ &  $0.00034$& $0.0096$& $0.0003$ \\
  \hline
 \end{tabular}
 \end{table*}

First, we notice that the ``tidal feedback" coefficient $Y$ can indeed be much larger than 1. For the Moon,
this coefficient amounts to $\sim 5$, which would imply a dominating $-\cos{\cal M}$ term in angle libration per
Eq. \ref{ans.eq}, and a much damped triaxiality driven libration. This, however, is not the case, as both
observational data (Table \ref{harm.tab}) and numerical simulations (Fig. \ref{moon.fig}) indicate. The reason
is explained in Section \ref{3.sec}. The dominant term is obviously $-\sin{\cal M}$, and it is still a triaxiality
driven spectrum, because the crucial condition \ref{zen.eq} is met. The presence of cosine-terms is only
betrayed by the asymmetry of the curve in Fig. \ref{moon.fig}, which is flatter at the top and sharper at the bottom,
indicating that the Moon spends a longer part of the orbital period at a spin rate above the resonance. But for the less 
viscous Io, condition \ref{cri.eq} is satisfied, and the libration curve in angle is almost exactly a cosine (Fig. \ref{io.fig}). We conclude that even among the bodies of the Solar system, a large variety of libration spectra should
be found, with relatively inviscid, icy satellites such as Io and possibly Titan exhibiting tidally driven,
$\cos{\cal M}$-dominated, libration of much reduced amplitudes due to the tidal feedback.

The case of Kepler 10b stands out because of subtle circumstances. On the one hand, $Y$ is much smaller than unity,
so the main term in the spectrum of angle libration is $-\sin{\cal M}$, just like the Moon. One of the reasons
for such small $Y$ is a low effective rigidity $\Lambda_2$ typical of massive super-Earth exoplanets.
This is not a triaxiality
driven spectrum, on the other hand, because $X\gg Z$, thus the the main action comes from the oscillating component
of the tidal torque. Although neither eccentricity nor $(B-A)/C$ are known for this exoplanets, and the numbers
given in Table \ref{moon.tab} are rather placeholders, the permanent figure of this massive planet is expected to be small,
and the eccentricity low too because of the high rate of tidal dissipation \citep{me14}. Therefore, the character of
libration is likely to be similar to that of the Moon, but for an entirely different reason.

The spectrum of libration of synchronous bodies is important for the tidal energy dissipation (and thus, heating)
and the tidal orbital evolution. \citet{me14} surmised that forced libration can boost the energy dissipation
in resonance allowing the resonant component of the secular torque to do some friction work. This may be valid
for triaxiality driven libration, where the secular tidal torque is in counter-phase with instantaneous velocity.
But we have determined that Io's libration must be tidally driven, with the sine-harmonic severely damped, so that the
boost to energy dissipation is expected to be negligible.

\label{lastpage}

\end{document}